\documentclass[amsmath,amssymb,twocolumn,showpacs,superscriptaddress]{revtex4}%
\usepackage{amsmath}
\usepackage{xcolor}
\usepackage{graphicx}% Include figure files
\usepackage{dcolumn}% Align table columns on decimal point
\usepackage{bm}% bold math
\usepackage{float}
\usepackage{epsfig}% FOR CORRECT FIGURE CAPTIONS
\usepackage{graphicx}%
\usepackage{amsmath}%
\usepackage{amsfonts}%
\usepackage{amssymb}
\usepackage{natbib}
\usepackage{soul}
\usepackage{color}

\usepackage{amssymb}

\usepackage{graphicx}
\begin{document}
\title{Robust ultrashort light bullets in strongly twisted waveguide arrays}
\author{Carles Mili\'{a}n}
\email{carmien@upvnet.upv.es}
\affiliation{ICFO--Institut de Ciencies Fotoniques, The Barcelona Institute of Science and Technology, 08860 Castelldefels (Barcelona), Spain}
\affiliation{Institut Universitari de Matem\`{a}tica Pura i Aplicada, Universitat Polit\`{e}cnica de Val\`{e}ncia, 46022 (Val\`{e}ncia), Spain}
\author{Yaroslav V. Kartashov}
\affiliation{ICFO--Institut de Ciencies Fotoniques, The Barcelona Institute of Science and Technology, 08860 Castelldefels (Barcelona), Spain}
\affiliation{Institute of Spectroscopy, Russian Academy of Sciences, Troitsk, Moscow, 108840, Russia}
\author{Lluis Torner}
\affiliation{ICFO--Institut de Ciencies Fotoniques, The Barcelona Institute of Science and Technology, 08860 Castelldefels (Barcelona), Spain}
\affiliation{Universitat Polit\`{e}cnica de Catalunya, 08034 Barcelona, Spain}
\begin{abstract}
We introduce a new class of stable light bullets that form in twisted waveguide arrays pumped with ultrashort pulses, where twisting offers a powerful knob to tune the properties of localized states. We find that above a critical twist, three-dimensional wavepackets are unambiguously stabilized, with no minimum energy threshold. As a consequence, when the higher order perturbations that accompany ultrashort pulse propagation are at play, the bullets dynamically adjust and sweep along stable branches. Therefore, they are predicted to feature an unprecedented experimental robustness.
\end{abstract}
\maketitle
%%%%%%%%%%
%  INTRODUCTION %
%%%%%%%%%%
The introduction by Silberberg several decades ago of the concept of {\it light bullets\/} \cite{silb}, as self-trapped spatiotemporal wavepackets of light, triggered an intense research activity (see \cite{tornerR,mihREV} and references therein). However, to date their experimental observation in steady state form is still an essentially open challenge. Three-dimensional stable states were known to exist, even at that time, in several mathematical models \cite{krz}, including parametric mixing in quadratic nonlinear media \cite{KR}, and have been subsequently studied in media with saturable \cite{bulsat},  competing \cite{bulcomp1,bulcomp2}, and nonlocal \cite{bang,bulnonl1,bulnonl2} nonlinearities, as well as in dissipative systems \cite{brambilla,buldiss1,buldiss2,buldiss3,rosanov,rosanov2}. Stable light bullets were predicted to form in discrete \cite{aceves,acevesprl95} and continuous \cite{mihPRE,mihPRL} lattices too, even when carrying vorticity \cite{lebl1,lebl3, cheskprl03}. Experimentally, landmark advances were achieved when fundamental \cite{minPRL,minPRA} and weakly unstable vortex \cite{pertschVB} light bullets were observed to form, albeit in a transient regime, in photonic crystal fibers with periodic cores and, later, when nonlinearity-induced locking of relatively long pulses in different modes, resulting in the formation of spatiotemporal localized wavepackets, was observed in graded-index media \cite{bulmmode1}.

Nevertheless, despite the intense efforts conducted during the last two decades that have led to several different theoretical proposals for the existence of stable light bullets (see, also, the recent review \cite{kartNR}), their experimental observation over long distances remains elusive \cite{minPRL,minPRA,pertschVB}. One of the salient challenges arises from the presence of \textit{higher order effects\/} (HOEs) that may destroy the bullet states existing in reduced mathematical models, where HOEs are disregarded. In practice, however, HOEs are actually significant and thus play an important role in experiments conducted with the ultrashort (sub-picosecond) pump pulses that are required in order to generate enough group-velocity dispersion for a spatiotemporal state to form over many dispersion lengths. As a consequence, to date, the experimental formation of long-lived bullets has not been achieved. 

In this Letter we report on the existence of stable light bullets that are predicted to be observable as robust three-dimensional wavepackets for unprecedented propagation distances even in the presence of HOEs. We found such states in twisted waveguide arrays, otherwise known to support stable spatial solitons \cite{kartOL,cuevas,saka}. Here we limit ourselves to bullets at the corners of square arrays, where centrifugal effects, key to our prediction, are maximal \cite{kartOL}, but our results can be extended to triangular or hexagonal geometries. It is important to realize that the centrifugal effects lead to the qualitative modification of the bullet properties, contrasting with the properties of bullets at the interfaces of static arrays \cite{mihOE} (or, in other words, three-dimensional generalizations of corner solitons \cite{makris,silPR}). Twisting offers a unique tunability knob for the existence, stability and localization of three-dimensional states, which is not available for other types of array modulations \cite{modlatt1,modlatt2}. At low twist rates, we found two separate stability domains that, above a critical twist rate, merge into a single stability domain extending up to zero energy. Therefore, the energy threshold for stability vanishes, therefore self-trapped tree-dimensional states can remain stable even in the presence of continuous energy leakage due to small perturbations. This is in contrast to the abrupt bullet disintegration predicted to occur, e.g., in straight arrays when the energy carried by the bullet drops below the corresponding threshold. 
%%%%%%%
%  FIGURE 1 %
%%%%%%%
\begin{figure}
\begin{center}
\includegraphics[width=.46\textwidth]{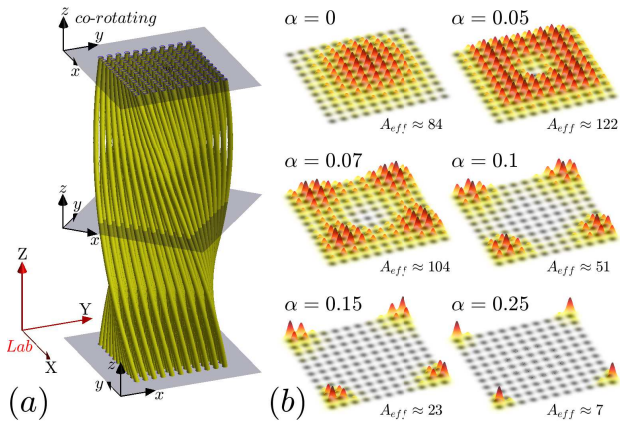}
\caption{(a) Twisted array of $11\times11$ waveguides. Axes are shown in the laboratory (red) and co-rotating (black) frames. (b) Amplitude profiles of the fundamental mode and effective mode area for different twist rates, $\alpha$. The grey background shows the potential cross section with $p=12$, $d=1.5$, $w=0.5$ (see definition of parameters below Eq.~\ref{eq1}). \label{f1}}
\end{center}
\end{figure}
%%%%%%%
%  FIGURE 2 %
%%%%%%%
%
\begin{figure*}
\begin{center}
\includegraphics[width=.92\textwidth]{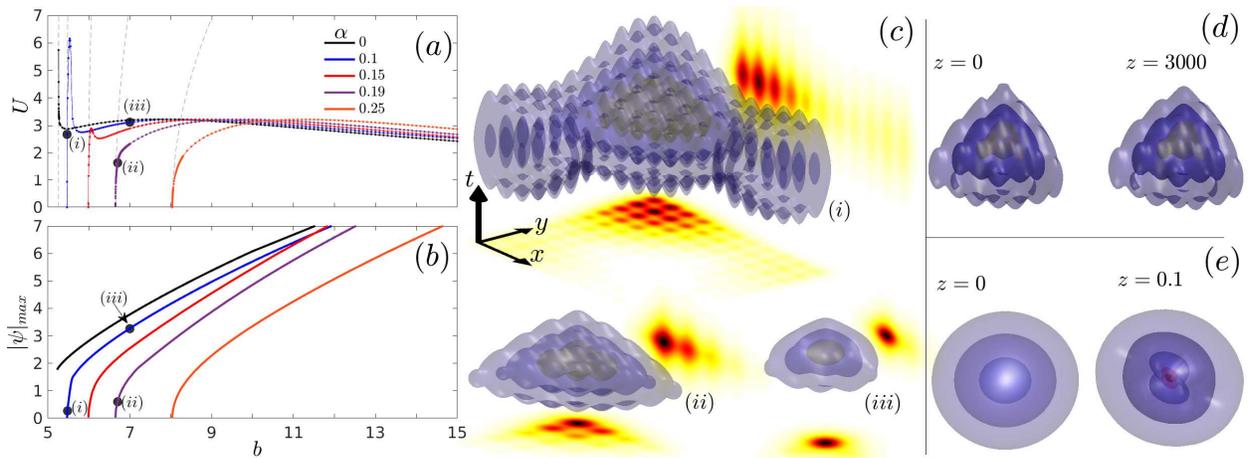}
\caption{(a) Energy and (b) peak amplitude of bullets vs propagation constant for several twist rates (see legend) in the array depicted in Fig.~\ref{f1}. (c) Stationary bullets corresponding to dots ($i$)-($iii$) in (a) and (b). The 2D projections show profiles in the $xt$ and $xy$ planes. (d), (e) show input (left), output (right) of propagation with Eq.~\ref{eq1} initiated with (d) stable [$\alpha=0.15$, $b=6.4$] and (e) unstable [$\alpha=0.15$, $b=12$] solutions. The 3D contours are taken at 1/10, 1/100, 1/1000 (c,d) and at 1/2, 1/4, 1/10, 1/20, 1/40 (e) of the maximum amplitude. The two inner red contours at $z=0.1$ in (e) show high intensity regions due to collapse at the bullet center.\label{f2}}
\end{center}
\end{figure*}
%
%%%%%%%
%  FIGURE 3 %
%%%%%%%
\begin{figure}
\begin{center}
\includegraphics[width=.46\textwidth]{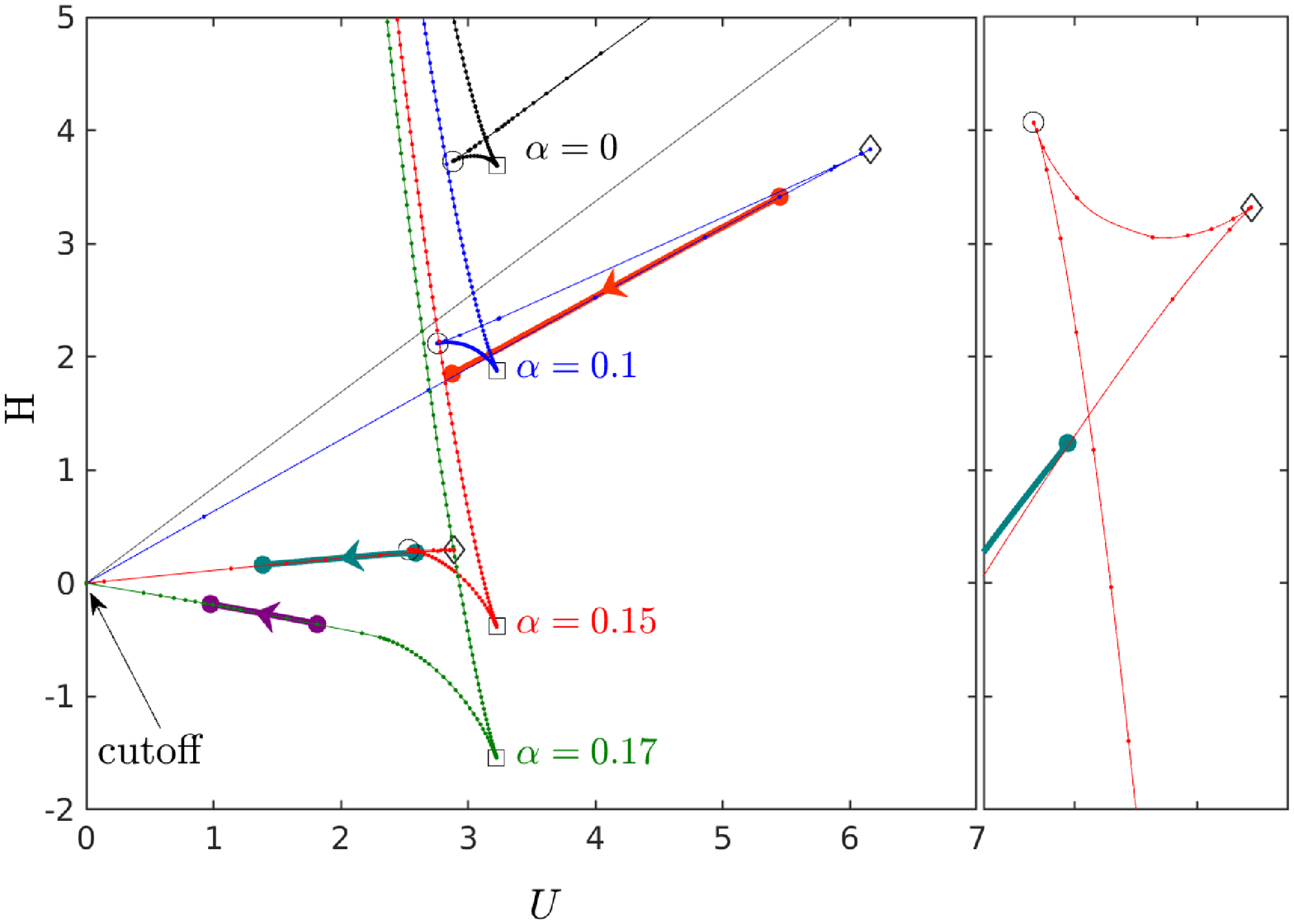}
\caption{Hamiltonian vs energy for bullet families with several values of $\alpha$. For visual purposes, we plot the quantity $\mathrm{H}\equiv H -$ constant x $U$ (with constant$=-6.1$), but the interpretations of  $\mathrm{H}(U)$ and $H(U)$ are equivalent. The right panel zooms around the swallowtail at $\alpha=0.15$ and cusps are highlighted with markers. The thick traces with arrows mark the dynamical evolution of the bullets under HOEs (see discussion after Eq.~\ref{eq2}). Input/output values are marked with filled circles.\label{f3}}
\end{center}
\end{figure}
%
%%%%%%%%%%
%%%%%%%%%%
%  MODEL %
%%%%%%%%%%

We address square waveguide arrays with constant twist rate around their center [c.f.\ Fig.~\ref{f1}(a)]. Such arrays can be realized in doped silica fibers \cite{ItanNC} where twists can be readily applied \cite{scienceRussel}. We analyze both truncated (without HOEs) and full (with HOEs) propagation regimes of spatiotemporal wavepackets along the $z-$axis. Under paraxial and slowly-varying envelope approximations, the evolution of linearly polarized light is governed by the dimensionless nonlinear Schr\"{o}dinger equation:
\begin{eqnarray}
i\partial_z \psi=-\frac{1}{2}\Delta\psi+i\alpha\left(x\partial_y-y\partial_x\right)\psi-(V+|\psi|^2)\psi-\mathcal{P}.\label{eq1}\end{eqnarray}
$\Delta\equiv\partial_x^2+\partial_y^2+\partial_t^2$, $x=[X\cos(\alpha z)+Y\sin(\alpha z)]/\mathrm{w}_0$ and $y=[Y\cos(\alpha z)-X\sin(\alpha z)]/\mathrm{w}_0$ are the scaled transverse coordinates in the co-rotating frame [Fig.~\ref{f1}(a)] with rate $\alpha\equiv2\pi/z_h$, $z\equiv Z/Z_R$ is the normalized propagation distance, $Z_R\equiv k_0n_0\mathrm{w}_0^2$ is the diffraction length, $\mathrm{w}_0$ is a reference width, $k_0=\omega_0/c$, $\omega_0$ is the carrier frequency, and $n_0$ is the unperturbed refractive index defining the dispersion $\kappa(\omega)\equiv n_0(\omega)\omega/c$, $t=[T-Z/v_g]/T_s$ is the time in the frame moving with group velocity $v_g$, $T_s\equiv\mathrm{w}_0\sqrt{-\kappa^{(2)}\kappa(\omega_0)}$ is the time scaling, $\kappa^{(2)}\equiv\partial_\omega^2\kappa(\omega_0)<0$ is the anomalous group velocity dispersion coefficient, $I=|\psi|^2/[k_0Z_Rn_2]$ is the intensity, and $n_2$ is the nonlinear index. For a square array of $N^2$ waveguides the potential $V=p\sum_{m,n=1}^N \exp\left\{-[(x-x_{m})^2+(y-y_{n})^2]/w^2\right\}$, where $(
x_{m},y_{n})$ are nodes of the grid with period $d$ and $p=\left(k_0\mathrm{w}_0\right)^2 n_0\delta n_\textrm{max}$ where $\delta n=n-n_0$. Our model accounts for radiation leakage in contrast to discrete models \cite{aceves,acevesprl95,mihOE}, particularly important for the dynamical tests of bullet robustness (c.f.,  Fig.~\ref{f5}), which may be affected by continuum modes. The function $\mathcal{P}$ accounts for the HOEs, described below in Eq.~\ref{eq2}.

An important feature of the system resides in its linear spectrum. By increasing the twist, the fundamental mode is expelled from the center [Fig.~\ref{f1}(b)] and localized around the corners due to the effective Coriolis force $\sim \alpha$ \cite{kartOL,kartOL2,saka}. Above $\alpha\approx0.05$ the effective mode area, $A_\textrm{eff}\equiv[\iint_{-\infty}^{+\infty}|\phi|^2 \textrm{d}x \textrm{d}y]^2 /\iint_{-\infty}^{+\infty}|\phi|^4\textrm{d}x \textrm{d}y$, decreases monotonically with $\alpha$ (see below).

We obtained three-dimensional stationary solutions of Eq.~\ref{eq1} without HOEs, $\psi=u(x,y,t)e^{ibz}$, using a modified square operator method \cite{yang}.
%%%%%%%%%%%%%%%%
% About FIG. 2 %
%%%%%%%%%%%%%%%%
Figure \ref{f2}(a) shows the energy $U\equiv\iiint_{-\infty}^{+\infty} |u|^2\mathrm{d}x\mathrm{d}y\mathrm{d}t$ vs $b$ for bullet families located at one corner of the array with various twist rates, $\alpha$. We checked that bullets located at the center of the array are almost insensitive to twist (not shown). Without twist ($\alpha=0$), families of bullets localized at the center or corner channels share the same qualitative features in terms of the $U(b)$ curves, hence comparison with \cite{mihPRE} is possible.
Stability of light bullets, checked via extensive propagation simulations [see Figs.~\ref{f2}(d,e)], coincides well with the Vakhitov-Kolokolov (VK) criterion \cite{VakhKol} predicting stability for $\partial_bU>0$ and instability otherwise. At low twist rates, $\alpha\lesssim0.17$, the $U(b)$ curves exhibit two stable regions. The stability domain on the right of the plot is equivalent to the stability domain reported for static arrays ($\alpha=0$) \cite{mihPRE,note1}. The stability domain on the left, at lower $b$ values, extends up to arbitrarily small energies $U\rightarrow0$ and terminates at the cut-off $b=b_\textrm{co}$. In this linear limit, bullet amplitudes $|\psi|_\textrm{max}$ vanish [Fig.~\ref{f2}(b)]. In the limit $\alpha\rightarrow0$, the $U(b)$ curve features a strong peak for $b \to b_\textrm{co}$, a reminiscence from free space \cite{mihPRE,kivbook,akhbook}. When $\alpha$ increases, such peak gradually disappears leading to the merging of the two stability domains into a single domain at $\alpha\approx0.17$.

In the limit $b\to b_\textrm{co}$, bullets asymptotically transform into $\psi=\phi(x,y)A(z,t)e^{i(b_\textrm{co}+\delta b)z}$, where $\phi$ is a linear mode of $V$, $A$ is the temporal envelope, and $\delta b$ the propagation constant offset from cutoff. Substituting the above ansatz into Eq.~\ref{eq1} and integrating the resulting equation for $A$, we get $U_\textrm{co}(\delta b)\approx\sqrt{8\delta b}A_\textrm{eff}$. In contrast to unbound systems, where $A_\textrm{eff}\rightarrow\infty$ causes $U_\textrm{co}$ to diverge \cite{kivbook,akhbook}, in the finite array $A_\textrm{eff}$ is finite and thus $U(b_\textrm{co})\equiv0$ as $\delta b\rightarrow0$. Moreover, a monotonic decrease of $A_\textrm{eff}$ with twist [see Fig.~\ref{f1}(b)] results in a considerable decrease of the slope $\partial_bU\vert_{b_\textrm{co}}=\sqrt{2/b}A_\textrm{eff}$ near the cutoff, which is related to the merging of stable regions. This asymptotic trend is marked in Fig.~\ref{f2}(a) by light-gray curves, in excellent agreement with the actual $U(b)$ curves around cutoff. We verified that the localized corner modes also form in twisted arrays with random fluctuations $\sim3\%$ \cite{itanOE,ItanNC} in channel positions and widths.

Figure \ref{f2}(c) shows light bullets for $\alpha=0.1$ [$(i)$, $(iii)$] and $\alpha=0.19$ [$(ii)$]. 
Such bullets have non-canonical shapes - they are strongly asymmetric in space because they are localized in the corners of the array due to interplay of centrifugal force and refraction, in contrast to previously reported symmetric states forming in the center of untwisted array. A new feature of such bullets is that in addition to nonlinearity, their localization degree is strongly affected by the twist rate of the array.
Note also that increasing the twist leads to stronger localization [c.f.\ bullets (i) and (ii)], as it happens when increasing $b$ [c.f.\ bullets (i) and (iii)]. We checked the stability of the bullet solutions by long ($z=3000$) three-dimensional propagation simulations initiated with stationary solutions perturbed with a $1\%$ noise in amplitude and phase. An illustrative stable evolution is shown in Fig.~\ref{f2}(d). Unstable states either decay or collapse. An example of the latter, that typically occurs for large $b$ values, is shown in Fig.~\ref{f2}(e), where high intensities develop around the bullet center.

%%%%%%%%%%%%%%%%%%%%%
% ABOUT H(U) [Fig. 3]
%%%%%%%%%%%%%%%%%%%%%
The transformation of the $U(b)$ curves with $\alpha$ [c.f.\ Fig.~\ref{f2}(a)] is reflected in the Hamiltonian-Energy diagrams shown in Fig.~\ref{f3}. The existence of cusps (see markers) corresponds to points where $\partial_bU=0$ [Fig.~\ref{f2}(a)]. When $\alpha$ increases, the cusps move in the $H(U)$ plane and two of them cease to exist for $\alpha\approx0.17$ when a single stability region forms [c.f.\ Fig.~\ref{f2}(a)].
In the $H(U)$ diagrams the bullet branches emerging from the cutoff correspond to straight lines departing from the origin with slope $b_\textrm{co}$. This is seen from the Hamiltonian of Eq.~\ref{eq1}, $H\equiv\iiint_{-\infty}^{\infty}\mathrm{d}x\mathrm{d}y\mathrm{d}t\mathcal{H}$, $\mathcal{H}=\mathrm{Re}(\psi^*[-\Delta/2+i\alpha\left(x\partial_y-y\partial_x\right)-V-|\psi|^2] \psi)+|\psi|^4/2$, which evaluated for a bullet reads $H=-bU+Q$ ($Q\equiv\iiint_{-\infty}^{\infty}\mathrm{d}x\mathrm{d}y\mathrm{d}t|\psi|^4/2$). Since $Q$ decreases faster than $U$ when $b\to b_\textrm{co}$, this limit yields $U,H\rightarrow0$ and $\partial_UH(b_\textrm{co})=b_\textrm{co}$.
For $\alpha=0$, the cutoff cannot be reached numerically \cite{note1} but the linear spectrum suggests that the cusp marked with a diamond is located at $U\sim100$, well outside the depicted area. Figure 3 depicts that twist impacts the shapes of the $H(U)$ diagrams strongly. Such shapes are important because their cusps and minima reveal stability, instability, and meta-stability, as in many other systems \cite{kusmartsev,akhHU,bang}.

\begin{figure}
\begin{center}
\includegraphics[width=.46\textwidth]{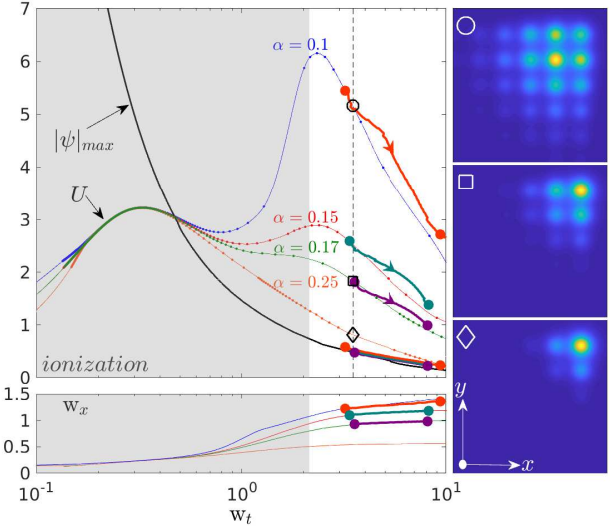}
\caption{Energy, amplitude, and spatial width of light bullet families versus temporal width. The $|\psi|_\textrm{max}(w_t)$ curves are very close to each other and they are all well represented by the single black curve. The grey area ($\mathrm{w}_t< 2.1$) marks the estimated ionization region in fused silica at $\lambda_0=1.55\ \mu$m and scaling factor $T_s=12~\textrm{fs}$. The insets show the bullet spatial cross sections with $\mathrm{w}_t=3.54$ for $\alpha=0.1$, $0.17$, $0.25$ (see markers in the main figure). The thick traces mark the evolution of $U$, $|\psi|_\textrm{max}$, $\mathrm{w}_x$, and $\mathrm{w}_t$ under the action of HOEs, corresponding to the traces in Fig.~\ref{f3}.\label{f4}}
\end{center}
\end{figure}

\begin{figure}
\begin{center}
\includegraphics[width=.46\textwidth]{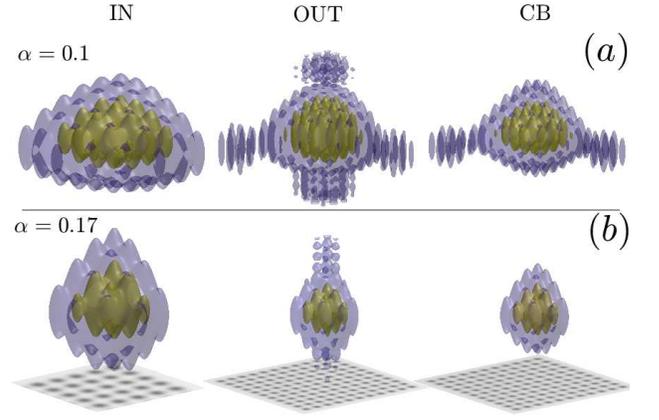}
\caption{Robust bullets with (a) $\alpha=0.1$, (b) $\alpha=0.17$. The $3D$ contours show input ($z=0$, left) and output ($z=300$, middle) profiles. The rightmost column shows the stationary \textit{corresponding bullet} (CB) to OUT profiles (see text). (a), (b) correspond to the orange, purple traces in Figs. \ref{f3}, \ref{f4}. A cross section of $V$ is shown for reference, bullets are at the top corner.\label{f5}}
\end{center}
\end{figure}

Experimentally observable properties of the bullets are illustrated in Fig.~\ref{f4}, which shows bullet families as peak amplitude $|\psi|_\textrm{max}$, energy $U$, and spatial width $\mathrm{w}_x$ vs temporal width $\mathrm{w}_t$: $\mathrm{w}_\xi\equiv[ \iiint_{-\infty}^{+\infty} (\xi-\xi_0)^2|u|^2\mathrm{d}x\mathrm{d}y\mathrm{d}t/U ]^{1/2}$, $\xi$ denotes $x$ or $t$, $\xi_0$ is evaluated at the bullet peak. The linear limit is at $\mathrm{w}_t\rightarrow\infty$, where $U, |\psi|_\textrm{max}\rightarrow0$. The VK stable regions correspond to $\partial_{\mathrm{w}_t} U<0$. 

%
%%%%%%%
%  FIGURE 4 %
%%%%%%%
%
%%%%%%%%%%%%%%%%%%%%%%%%%%%%%%%%
% Higher Order Effects  [Fig. 5]
%%%%%%%%%%%%%%%%%%%%%%%%%%%%%%%%

Next we address the robustness of the bullets in the presence of HOEs that enter into play in material systems pumped with sub-picosecond light pulses. In fused silica, relevant HOEs are:
\begin{align}
&\mathcal{P}\approx\frac{1}{|\kappa^{(2)}|} \sum_{q=3}^{\infty}\frac{\kappa^{(q)}}{T_s^{q-2}}\frac{(i\partial_t)^q}{q!}\psi+\frac{i\partial_\tau(|\psi|^2\psi)}{\omega_0T_s}-\label{eq2}\\ &
\nonumber-\mu\psi\left(|\psi|^2-\int_{-\infty}^\infty\mathrm{d}\tau'|\psi(\tau-\tau')|^2h_R(\tau')\right)+i\frac{b_K}{2}|\psi|^{2K-2}\psi,
\end{align}
and account, respectively, for higher-order dispersion \cite{silica}, self-steepening, Raman scattering, and multi-photon absorption (MPA): $\kappa^{(q)}\equiv\partial_\omega^q\kappa(\omega_0)$, $h_R(\tau)=(\tau_1^2+\tau_2^2)/(\tau_1\tau_2^2)\theta(\tau)\exp{-\tau/\tau_2}\sin(\tau/\tau_1)$, $\tau_1=12.2(\textrm{fs})/T_s$, $\tau_2=32(\textrm{fs})/T_s$, $\theta(\tau)$ is the Heaviside function, $b_K\equiv\beta_KZ_R/[k_0n_2Z_R]^{k-1}$, and $K\equiv \lfloor\hbar\omega_0/U_i+1\rfloor=12$ ($\lfloor \cdot\rfloor$ denotes the floor function) is the number of simultaneously absorbed photons at the wavelength $\lambda_0\sim1.55\ \mu$m, where $U_i\approx9$ eV is the ionization potential \cite{couPRB}. Also, $\beta_K=7.3\times10^{-169}\ m^{21}/W^{11}$ is the MPA coefficient, which accounts for ionization in the absence of impurities and light intensities well below the tunnel ionization regime (see, e.g., \cite{couPR}). Obviously, we consider bullets at intensity levels well below the filamentation onset, thus MPA is negligible with respect to the Kerr effect, i.e., $[b_K|\psi|^{2K-2}/2]/|\psi|^2\lesssim10^{-5}$ \cite{couPRB,couPR,milianjosab,minapl}. For $\textrm{w}_0=30~\mu \textrm{m}$ the temporal scaling is $T_s=12~\textrm{fs}$, the index contrast is $\delta n\approx6\times10^{-4}$ $(p=12)$ in Ge-doped glass \cite{prismyan,note2,butov}, $Z_R\approx5$ mm, $Z_h\approx30$ cm (for $\alpha=0.1$).

To elucidate the robustness of the bullets we propagated them numerically in the presence of all HOEs included in Eq.~\ref{eq2} over the distance $z=300\approx1.55$ m$\approx 30Z_{d}$, where $Z_{d}\equiv T_s^2\mathrm{w}_t^2/|\kappa^{(2)}|$ is the dispersion length and $\kappa^{(2)}\approx28$ fs$^2/$mm. Illustrative input (IN) and output (OUT) profiles are shown in Fig.~\ref{f5}, in the left and middle columns. With the above scaling, input bullets have peak intensities of the order of $40$ GW/cm$^2$. The corresponding variation of the bullet parameters ($H$, $U$, $|\psi|_\textrm{max}$ $\mathrm{w}_x$, $\mathrm{w}_t$) upon propagation is shown by the thick traces with arrows in Figs.~\ref{f3} and \ref{f4}. Over such propagation distances, bullets broaden from $\mathrm{w}_t\approx43$ fs to $\approx120$ fs. The traces show that the values of all parameters follow well those of the stationary bullet families, i.e., as HOEs are perturbative they smoothly transform a given bullet into a new one of the same stable family with lower $U$. The sweep along a family occurs because Raman and self-steepening break the global conservation of $H$ and cause the fraction of total $H$ and $U$ carried by the bullets to vary \cite{akh96,skryRMP}, but solutions remain close to the stable $H(U)$ branch (Fig.~\ref{f3}).
The emitted radiation is clearly seen in the output (OUT) profiles in Fig.~\ref{f5} [middle column]. For completeness, we have added the right column showing the profile of the ideal stationary bullet with the same parameters as the output state (OUT). These profiles are termed \textit{corresponding bullet} (CB) and are identical to those at the output (OUT), apart from the weak radiation.
%%%%%%%
%  FIGURE 5 %
%%%%%%%

In summary, we highlight that twisting fiber arrays affords a powerful physical mechanism to impact the properties of multi-dimensional wavepackets, which we found to result in {\it stable three-dimensional light bullets without energy threshold\/} that exist, {\it above a critical twist\/}, as robust states localized at the corners of the array. Our results suggest that long-lived propagation of such light bullets should be experimentally feasible with ultrashort pump pulses and moderate peak powers. With a suitable design of the array and pump conditions, the bullet energy, peak amplitude and widths (in space and time) can remain close to the values corresponding to the stationary solutions as bullet evolve in the presence of HOEs. Dynamical evolutions where stable multidimensional solitons are stationary in the presence of HOEs was reported previously only in cavities (see, e.g., \cite{brambilla,milianPRL}). Here we found that a similar phenomenon occurs in single-pass twisted arrays. Our calculations were performed for the parameters of structured Ge-doped silica fibers that can be fabricated with lengths far exceeding the twist periods $\sim30$ cm, with diffraction $\sim k_0n_0\mathrm{w}_0^2 \mathrm{w}_x^2$ and dispersion $\sim T_s^2\mathrm{w}_t^2/|\kappa^{(2)}|$ lengths of the order of centimeters, for pump pulses with FWHM $\sim 50$ fs. We assumed HOEs to remain perturbative, an assumption that we verified can hold in Ge-doped silica fibers under properly-designed experimental conditions. 
\begin{acknowledgments}
{\small This paper is dedicated to the memory of Dr Yaron Silberberg, who passed away recently. The work was supported by the Government of Spain through Juan de la Cierva Incorporaci\'{o}n, Severo Ochoa SEV-2015-0522, and FIS2015-71559-P; Generalitat de Catalunya; CERCA; Fundaci\'{o} Cellex; and Fundaci\'{o} Mir-Puig. CM thanks Profs. J. Alberto Conejero Casares and Pedro Fern\'{a}ndez de C\'{o}rdoba Castell\'{a} for their support.}
\end{acknowledgments}
%
%%%%%%%%%
% BIBLIO
%%%%%%%%%


\begin{thebibliography}{10}
\bibitem{note1}
The threshold-less stability regions shown in Fig. \ref{f2}(a) are extremely sharply peaked for small twist values $\alpha\ll0.1$, and they become technically difficult to be numerically explored.
This is presumably why this very relevant feature, in principle general for any truncated system, has not been reported in the previous numerical studies. In the current work, our numerical method was incapable of tracing the $U(b)$ down to the cutoff for $\alpha=0$.
\bibitem{note2}
Ge-doped silica is envisaged as a promising platform to make the arrays because the index contrast, $\Delta n$, related to the potential depth, $p$, increases slightly with wavelength and the existence of stable bullets is guaranteed, contrarily to the case of F-doped glass \cite{butov}. In the latter case, $p$ is reduced as bullet's red-shift and hence existence of stable bullets becomes compromised \cite{mihPRE}.
\end{thebibliography}
\end{document}